\documentclass[pra,tightenlines,showpacs,letterpaper, preprintnumbers,nofootinbib,floatfix]{revtex4}

\usepackage{amsmath,amssymb,amsfonts,graphicx,pifont,dcolumn}
\usepackage{epstopdf}
\usepackage{color}
\usepackage{cancel}
\usepackage{ulem}
\usepackage{xfrac}
\RequirePackage{slashed}
\begin{document}

\title{Relation between the Berry phase in quantum hermitian and non-hermitian systems and the Hannay phase in the equivalent classical systems.}

\author{H.~Fanchiotti} \author{C.A.~Garc\'\i a Canal}
\affiliation{IFLP(CONICET) and Departamento de F{\'\i}sica, Universidad  Nacional de La Plata,  C.C. 67 1900, La Plata, Argentina}

\author{ M.~Mayosky} 
 \affiliation{LEICI, Departamento de Electrotecnia Facultad de Ingenier{\'\i}a, Universidad Nacional de La Plata, La Plata,
Argentina and  Comisi\'on de Investigaciones Cient{\'\i}ficas de la Provincia de Buenos Aires-CICpBA, Argentina}

 \author{A.~Veiga}
\affiliation{LEICI, Departamento de Electrotecnia Facultad de Ingenier{\'\i}a, Universidad Nacional de La Plata, La Plata, Argentina; CONICET}

\author{V.~Vento}
\affiliation{Departamento de F{\'\i}sica Te\'orica-IFIC. Universidad de Valencia-CSIC, E-46100, Burjassot (Valencia), Spain}

\date{\today}%

\begin{abstract}

The well-known geometric phase present in the quantum adiabatic evolution discovered by Berry many years ago has its analogue, the Hannay phase, in the classical domain.
We calculate the Berry phase with examples  for quantum hermitian and non-hermitian $PT$-symmetric Hamiltonians and compare with the Hannay phase in their classical equivalents.
We use the analogy to propose resonant electric circuits which reproduce the theoretical solutions in simulated laboratory experiments.
\end{abstract}


\maketitle

 \section{Introduction}
In a recent paper \cite{AJP} the measurement of the geometric Hannay angle \cite{Hannay}
was studied in connection with the Foucault pendulum and an analog electric circuit. This analysis
induced us to find the explicit connection between the Hannay phase and the well known
quantum Berry phase \cite{Berry} on the basis of the mathematical equivalence
between classical and quantum dynamics presented in ref. \cite{AP}.  In here we generalize the relation between the two phases for several interesting quantum systems and describe the analog electric circuits. For that purpose we start by recalling in section \ref{classical} the mathematical formalism of
decomplexification  \cite{Arnold} and the construction of equivalent classical equations of motion for effective hamiltonians  \cite{AP}.  In section III we calculate the Berry phase in
a quantum adiabatic evolution process for hermitian Hamiltonians by solving exactly the Schr\"odinger equation and compare the result with previous work with classical Hamiltonians leading to the Hannay phase. We find explicitly the well known relation between the Berry phase and the classical Hannay angle \cite{Berry1}. We generalize  the calculation and the comparison to PT-symmetric quantum non-hermitian Hamiltonians. In Section IV  we establish the correspondence of the quantum computation with the equivalent classical analogs which lead to equations which are similar to those of  resonant electric circuits coupled by gyrators. These results lead us to propose and simulate laboratory experiments for measuring the equivalent geometric phases in section V. We draw some conclusions of our study in the final section.

\section{Classical analogs of quantum systems}
\label{classical}

We describe next  the decomplexification formalism which defines a classical analog of a quantum system in order to  establish a connection between quantum and classical dynamics, which result ultimately in an analog description of the systems in terms of electric circuits. This analogy leads to feasible laboratory experiments which helps one understand complex concepts like geometric phases.

\subsection{Decomplexification of quantum Hamiltonians}

Let us recall  the decomplexification procedure \cite{Arnold} of
quantum Hamiltonians. The quantum evolution of a system is governed by the Schr\"odinger equation

\begin{equation}
i\, \frac{\partial }{\partial t}\Psi = H\,\Psi.
\end{equation}
We use $\hbar=1$ unless a quantum effect needs to be emphasized. For a two component system the Hamiltonian $H$ is a $2 \times 2$ matrix
with matrix elements that in general are complex and eventually time-dependent. Moreover, $H$ can be a hermitian or a non-hermitian
operator treated as an effective Hamiltonian. $\Psi$ in this case is a two component vector.

Let us write explicitly  the Schr\"odinger  equation in terms of real and imaginary parts of both $\Psi$ and $H$,
\[
\Psi =\left[
\begin{array}{c}
\Psi _{1} \\
\Psi _{2}%
\end{array}%
\right] =\left[
\begin{array}{c}
Re(\Psi _{1}) \\
Re(\Psi _{2})
\end{array}%
\right] +i\left[
\begin{array}{c}
Im(\Psi _{1}) \\
Im(\Psi _{2})
\end{array}%
\right]
\]
and
\[
 H=\left[\begin{array}{cc}
H_{11} & H_{12} \\
H_{21} & H_{22}
\end{array}\right]%
=\left[
\begin{array}{cc}
Re(H_{11}) & Re(H_{12}) \\
Re(H_{21}) & Re(H_{22})
\end{array}%
\right] +i\left[
\begin{array}{cc}
Im(H_{11}) & Im(H_{12}) \\
Im(H_{21}) & Im(H_{22})
\end{array}%
\right]
\]
In order to simplify the notation, we introduce
\[
Re(\Psi _{i})=x_{i}\,\,\,;\,\,\,Im(\Psi _{i})=y_{i}\,\, ;\, i=1,2\, ,
\]
\[
Re(H_{ij})=f_{ij}\,\,\,;\,\,\,Im(H _{ij})=g_{ij}\,\, ; \,i, j = 1,2\, .
\]
Separating real and imaginary parts we get
\begin{equation}
\left[
\begin{array}{c}
\dot{x}_{1} \\
\dot{x}_{2}%
\end{array}%
\right] =\left[
\begin{array}{cc}
g_{11} & g_{12} \\
g_{21} & g_{22}
\end{array}%
\right] \left[
\begin{array}{c}
x_{1} \\
x_{2}%
\end{array}%
\right] +\left[
\begin{array}{cc}
f_{11} & f_{12} \\
f_{21} & f_{22}
\end{array}%
\right] \left[
\begin{array}{c}
y_{1} \\
y_{2}%
\end{array}%
\right]
\label{xdot}
\end{equation}
and
\begin{equation}
\left[
\begin{array}{c}
\dot{y}_{1} \\
\dot{y}_{2}%
\end{array}%
\right] =\left[
\begin{array}{cc}
g_{11} & g_{12} \\
g_{21}) & g_{22}
\end{array}%
\right] \left[
\begin{array}{c}
y_{1} \\
y_{2}%
\end{array}%
\right] -\left[
\begin{array}{cc}
f_{11} & f_{12}\\
f_{21} & f_{22}
\end{array}%
\right] \left[
\begin{array}{c}
x_{1} \\
x_{2}%
\end{array}%
\right].
\end{equation}
Using matrix notation we can write these equations as,
\begin{equation}
\left[
\begin{array}{c}
\dot{x}_{1} \\
\dot{y}_{1} \\
\dot{x}_{2} \\
\dot{y}_{2}%
\end{array}%
\right] =\left[
\begin{array}{cccc}
\;\; \,g_{11} & f_{11} &\;\;\, g_{12} & f_{12} \\
- f_{11} & g_{11} & - f_{12} & g_{12} \\
\;\;\, g_{21} & f_{21} &\;\; \,g_{22} & f_{22} \\
- f_{21} & g_{21} & - f_{22} & g_{22}
\end{array}%
\right] \left[
\begin{array}{c}
x_{1} \\
y_{1} \\
x_{2} \\
y_{2}%
\end{array}%
\right].
\label{ydot}
\end{equation}
This result coincides with the general case presented in ref. \cite{Arnold} when restricted to two dimensions.

Let us introduce further simplifying notation
\[
A=\left[
\begin{array}{cc}
g_{11} & g_{12} \\
g_{21} & g_{22}
\end{array}%
\right] \,\,\,\,\,\,;\,\,\,\,\,\,B=\left[
\begin{array}{cc}
f_{11} & f_{12}\\
f_{21} & f_{22}
\end{array}%
\right]
\]
and
\[
x=\left[
\begin{array}{c}
x_{1} \\
x_{2}%
\end{array}%
\right]  \,\,\,\,\,\,;\,\,\,\,\,\,y=\left[
\begin{array}{c}
y_{1} \\
y_{2}%
\end{array}%
\right]
\]
Consequently, the first order differential equations Eqs.( \ref{xdot}) and (\ref{ydot}) become
\begin{eqnarray}
\dot{x}&=&A\,x+B\,y \nonumber \\
\dot{y}&=&A\,y-B\,x \nonumber
\end{eqnarray}

These equations can be easily transformed into two separated second order differential equations for $x$ and $y$, namely

\begin{eqnarray}
\label{coupled}
\ddot{x}&=&(A+B\,A\,B^{-1})\,\dot{x} - (B\,A\,B^{-1}\,A\,+B\,B)\,x , \nonumber \\
\ddot{y}&=&(A-B\,A\,B^{-1})\,\dot{y} +(B\,A\,B^{-1}\,A\,- B\,B)\,y.
\end{eqnarray}

From the mathematical point of view, we have transformed the quantum evolution equation into
a pair of equations for coupled harmonic oscillators. Certainly, the above procedure can be
immediately extended to the case of a Hamiltonian of finite arbitrary
dimension. The connection between the Schr\"odinger equation and
harmonic oscillators has been obtained also by using different techniques \cite{others}.

\subsection{Effective Hamiltonian}

Let us apply the formalism to the standard expression for an effective Hamiltonian of the form

\begin{equation}
H=M-i\,\Gamma,
\end{equation}
where for two degrees of freedom $M$ and  $\Gamma $
are $2 \times 2$ hermitian matrices.  These matrices can be diagonalized and their corresponding eigenvalues are real.

We are interested in the behavior of the probability densities in different situations associated with the form of $M$ and of $\Gamma$.
From the Schr\"odinger equation and its adjoint, one immediately finds that the modulus of the state function evolves in time according to
\[
\frac{d}{dt}\left( \Psi
^{\dagger }\Psi \right) =-2\Psi ^{\dagger }\Gamma \Psi.
\]
This expression shows that the probability density is controlled by $\Gamma$., i.e., the probability will stay constant or will decay in time according to the
structure of the $\Gamma$-matrix. 

The characteristic polynomial
of this matrix is given by
\[
P(\lambda )=\lambda ^{2}-Tr(\Gamma )\lambda +\det (\Gamma )
\]
thus, the evolution of the probability density, driven by the eigenvalues of $\Gamma$, depends
on the trace and the determinant of the $\Gamma$ matrix.

Let us study first the case for which $Tr(\Gamma ) >0$ and $\det (\Gamma ) > 0$ which imply that
its eigenvalues are positive. Then, the most simple form of the matrix is
\begin{equation}
\Gamma =\left[
\begin{array}{cc}
s & 0 \\
0 & s%
\end{array}%
\right],
\end{equation}
and with this shape of $\Gamma $ the probability density will decrease in time.

The other possibility is when $Tr(\Gamma )=0$ and $\det (\Gamma) < 0$. Consequently, the simple form
of $\Gamma $ reads

\begin{equation}
\Gamma =\left[
\begin{array}{cc}
-s & 0 \\
0 & s
\end{array}
\right].
\end{equation}
This particular form implies that the evolution is given by
\[
\frac{d}{dt}\left( \Psi ^{\dagger }\Psi
\right) =-2\,s\left( \Psi _{1}^{\ast }\Psi _{1}-\Psi _{2}^{\ast }\Psi
_{2}\right).
\]
Here, differently from the previous case, the term on the right has not a definite sign,
because we can also write a  $\Gamma$-matrix with the same properties with the positions of
$s$ and of $-s$ exchanged,

\begin{equation}
\Gamma =\left[
\begin{array}{cc}
s & 0 \\
0 & -s
\end{array}
\right],
\end{equation}
with the result,

\[
\frac{d}{dt}\left( \Psi ^{\dagger }\Psi
\right) = 2\,s\left( \Psi _{1}^{\ast }\Psi _{1}-\Psi _{2}^{\ast }\Psi
_{2}\right).
\]
Consequently,  both equalities can only be satisfied if
\[
\Psi_{1}^{\ast}\Psi_{1}-\Psi_{2}^{\ast}\Psi_{2}=0.
\]
Thus in this case  the probability density is conserved, even if the Hamiltonian has the imaginary part $\Gamma$,
certainly of this very special structure. This kind of Hamiltonian is an example of the well-known $PT$-symmetric
non-hermitian Hamiltonian \cite{PT} that we shall discuss in detail below.

We continue with the analysis considering a hermitian two-dimensional Hamiltonian with equal diagonal elements
\begin{equation}
H=\left[
\begin{array}
[c]{cc}%
h& f-ig\\
f+ig & h
\end{array}
\right]
\end{equation}
then, the matrices $A$ and $B$ above are simply
\[
A=\left[
\begin{array}
[c]{cc}%
0 & -g\\
g & 0
\end{array}
\right]\,\,\,\,\,;\,\,\,\,\,B=\left[
\begin{array}
[c]{cc}%
h & f\\
f & h
\end{array}
\right].
\]
These matrices define the equivalent coupled harmonic oscillator problem Eq.(\ref{coupled}).

Let us discuss some examples:

\noindent i) The case: $g=0$.

This case is related with the classical analog of two resonant electric circuits coupled inductively
\cite{Schindler}. It has also a mechanical analog consisting of two pendulums coupled by means of a
spring. The analogy is easily seen from the corresponding values of the $A$ and $B$ matrices,

\[
A=\left[
\begin{array}
[c]{cc}%
0 & 0\\
0 & 0
\end{array}
\right]\,\,\,\,\,;\,\,\,\,\,B=\left[
\begin{array}
[c]{cc}%
h & f\\
f & h
\end{array}
\right].
\]

\noindent ii) The case $f=0$.

This case has the classical analog of two resonant electric circuits coupled by means of a gyrator.
Remember that a gyrator is a non-reciprocal passive
element of two ports \cite{Bala} with a given conductance.
In this case, again the relation is fixed by the corresponding $A$ and $B$ matrices,

\[
A=\left[
\begin{array}
[c]{cc}%
0 & -g\\
g & 0
\end{array}
\right]\,\,\,\,\,;\,\,\,\,\,B=\left[
\begin{array}
[c]{cc}%
h & 0\\
0 & h
\end{array}
\right].
\]

In both  cases, the  differential equations are written in terms of the state variables, voltage
in each partial circuit. It is worth mentioning that this circuit network is also
related to a dimer, an oligomer consisting of two monomers joined by bonds \cite{dimer}.
When the coupling is inductive, it appears directly related to the voltage variable, while for the
case of the gyrator, the coupling is connected to the first derivative of voltages in the differential equations.
This last case was analyzed in \cite{AJP} in connection with the Foucault pendulum.

\section{Berry and Hannay geometrical phases}

We will come back to the relation between the quantum systems and the classical analog of resonant electric circuits but next we proceed to  study the Berry phase, the geometric phase associated to the adiabatic quantum evolution of our effective Hamiltonian, by solving the Schr\"odinger equation, and will compare it with the Hannay phase, the classical geometric phase analog \cite{Hannay}.

\subsection{Hermitian Hamiltonian}

In conventional Hilbert spaces associated to quantum dynamical systems driven by a hermitian Hamiltonian, a general two-dimensional  quantum state can be written as

\begin{equation}\label{gs}
\left|\Psi(t)\right\rangle=a(t)\left|  1\right\rangle +b(t)\left|  2\right\rangle,
\end{equation}
where $\left|  1\right\rangle $ and $\left|  2\right\rangle $ stand for the eigenstates of  $H$
with eigenvalues $\lambda_{1}$ and $\lambda_{2}$ respectively. The
initial state at $t=0$, when expressed on the Bloch sphere of the Hilbert space reads
\[
\left|\Psi(0)\right\rangle=\cos\frac{\theta_{0}}{2}\left|  1\right\rangle +\sin\frac{\theta_{0}%
}{2}\,e^{i\varphi_{0}}\left|  2\right\rangle.
\]
We now replace in the  time dependent Schr\"odinger equation
\[
i\frac{d}{dt}\left|\Psi(t)\right\rangle=H\,\left|\Psi(t)\right\rangle,
\]
the general solution (\ref{gs}) and get the following differential equations for
$a(t)$ and $b(t)$
\[
i\dot{a}(t)=\lambda_{1}a(t)\,\,\,\,\,;\,\,\,\,\,i\dot{b}(t)=\lambda_{2}b(t),
\]
whose solutions are
\[
a(t)=\alpha\,e^{-i\lambda_{1}\,t}\,\,\,\,\,;\,\,\,\,\,b(t)=\beta\,e^{-i\lambda_{2}\,t}.
\]
Thus the general solution results
\[
\left|\Psi(t)\right\rangle=\alpha\,e^{-i\lambda_{1}\,t}\left|  1\right\rangle +\beta\,e^{-i\lambda_{2}%
\,t}\left|  2\right\rangle
\]
where the parameters $\alpha$ and $\beta$ are fixed by the initial conditions
and  become
\[
\alpha=\cos\frac{\theta_{0}}{2}\,\,\,\,\,;\,\,\,\,\,\beta=\sin\frac{\theta_{0}}
{2}\,e^{i\varphi_{0}}.
\]
Finally, the general solution is
\[
\left|\Psi(t)\right\rangle=\cos\frac{\theta_{0}}{2}\,e^{-\lambda_{1}\,t}\left|  1\right\rangle
+\sin\frac{\theta_{0}}{2}\,e^{i\varphi_{0}}e^{-i\lambda_{2}\,t}\left|
2\right\rangle,
\]
that we write as
\begin{equation}
\left|\Psi(t)\right\rangle=e^{-\lambda_{1}i\,t}\left(  \cos\frac{\theta_{0}}{2}\left|
1\right\rangle +\sin\frac{\theta_{0}}{2}\,e^{i\varphi_{0}}e^{i\,t(\lambda
_{1}-\lambda_{2})}\left|  2\right\rangle \right).
\end{equation}

From this expression we can determine that the system under consideration will return to the initial state after a time $T$
given by
\[
T\,(\lambda_{1}-\lambda_{2})=2\,\pi,
\]
when the state takes the form
\[
\left|\Psi(T)\right\rangle=e^{-i\lambda_{1}\,T}\left(\cos\frac{\theta_{0}}{2}\left|  1\right\rangle
+\sin\frac{\theta_{0}}{2}\,e^{i\varphi_{0}}\left|  2\right\rangle
\right)=e^{-i\lambda_{1}\,T}\left|\Psi(0)\right\rangle.
\]

This expression shows that after the evolution, the quantum system
returns to the initial state but having acquired a total extra phase
\begin{equation}
\varphi_{total}=-\lambda_{1}\,T=-2\,\pi\frac{\lambda_{1}}{(\lambda_{1}-\lambda_{2})}.%
\label{total}
\end{equation}
This total phase includes besides the standard dynamical phase, another one, the Berry phase which
is of geometrical origin. In order to detect this geometrical phase we compute
the dynamical phase associated with the adiabatic motion
\[
\varphi_{dynamic}=-\int_{0}^{T}\left\langle \Psi(t)\right|  H\left|
\Psi(t)\right\rangle dt,
\]
giving rise to
\begin{equation}
\varphi_{dynamic}=-2\,\pi\,\frac{\lambda_{1}}{(\lambda_{1}-\lambda
_{2})}+2\,\pi\,\sin^{2}\frac{\theta_{0}}{2}.%
\label{dynamic}
\end{equation}
The Berry geometric phase \cite{Berry}, is obtained from Eqs. (\ref{total}) and (\ref{dynamic})

\begin{equation}
\varphi_{geometric}=-2\,\pi\,\frac{\lambda_{1}}{(\lambda_{1}%
-\lambda_{2})}-\varphi_{dynamic}=-2\,\pi\,\sin^{2}\frac{\theta_{0}}{2}=\pi,
(\cos\theta_{0}-1)
\label{factor2}
\end{equation}
where its geometric origin is evident since it is independent of the dynamics determined by $H$.
Moreover, this Berry phase coincides, except for a factor $2$ and a change of sign,
with the Hannay phase \cite{Hannay} of the classical equivalent system, the Foucault pendulum, as it was determined
in \cite{AJP},

\begin{equation}
\triangle\phi = 2\,\pi - \eta = 2\,\pi\,(1- \sin(\lambda)) = 2\,\pi\,(1- \cos(\theta))
\label{phaserelation}
\end{equation}
where the phase is written in terms of the colatitude $\theta$ as it is usually presented.

Eq.(\ref{phaserelation}) explicitly shows the connection between the classical Hannay angle and the quantum Berry phase, namely

\begin{equation}
\triangle\phi = -2 \varphi_{geometric}.
\label{berryrelation}
\end{equation}
This is a particular case of the general result \cite{Berry1,BH} 

\begin{equation}
\triangle\phi_l=\frac{\partial \varphi_{Hannay}}{\partial n_l}.
\end{equation}
The factor 2 in Eq.(\ref{berryrelation}) comes  from
the quantum vacuum contribution to the energy which in the harmonic oscillator is
$E_n \sim (n+\frac{1}{2})\hbar \omega$ \cite{Berry1}.

\subsection{Non-hermitian $PT$-symmetric  Hamiltonian}

The basic property of the operators that represent observables
in Quantum Mechanics is hermiticity, because these operators
have real eigenvalues that can be considered measurable
quantities related to the corresponding physical magnitudes. Whenever open systems exhibit flows
of energy, particles and information, they are described by 
non-hermitian Hamiltonians, in general associated with the
decay of the norm of a quantum state. Among the non-hermitian
Hamiltonians, those that obey parity-time ($PT$)
symmetry are of particular interest because they can admit
real eigenvalues while describing physical open systems which
present balanced loss into and gain from the surrounding
environment. Besides,
as a parameter, let us call it $\gamma$, that is associated with
the degree of non-hermiticity of $H$, changes, a spontaneous
$PT$-symmetry breaking occurs (see the Appendix) and the real properties of
eigenvalues are lost, they become complex \cite{PT}.

Let us study the two dimensional system characterized by

\begin{equation}
H= M -i\, \Gamma = \left[
\begin{array}
[c]{cc}%
a & -ig\\
ig & a
\end{array}
\right]  -i\left[
\begin{array}
[c]{cc}%
-s & 0\\
0 & s
\end{array}
\right]
\end{equation}
having eigenvalues
\[
a+\sqrt{\left(  -s^{2}+g^{2}\right)  }\,\,\,\,;\,\,\,\,
a-\sqrt{\left(  -s^{2}+g^{2}\right)  }
\]
that can be written as
\[
a+g\sqrt{\left(  -\gamma^{2}+1\right)  }\,\,\,\,;\,\,\,\,a-g\sqrt{\left(  -\gamma^{2}+1\right)  }
\]
with
\[
\gamma=\frac{s}{g}
\]
Notice that in the limit $\gamma\rightarrow 0$ the hamiltonian becomes hermitian.
If $\gamma^{2}\leq 1$, both eigenvalues are real and the $PT$-symmetry present
in the Hamiltonian is also present in the solutions. The symmetry is unbroken.
The particular value $\gamma^{2} = 1$, is known as an exceptional point where
the spontaneous breaking of the $PT$-symmetry appears and the eigenvalues from then on become complex conjugate quantities.

In order to clarify the analysis of the geometric phase and due to the fact that
the $M$ and $\Gamma$ components of $H$ do not commute, it is necessary to
use the biorthogonal quantum formalism \cite{Brody}.

Consider the symmetric situation. The dynamical equation is given by
\[
i\,\frac{\partial
}{\partial t}\left|\Psi\right\rangle=H\,\left|\Psi\right\rangle= (M-i\,\Gamma) \,\left|\Psi\right\rangle
\]
with the initial condition
\[
\left|  \Psi\left(  0\right)  \right\rangle
=\cos\left(  \frac{\theta}{2}\right)  \left|  \phi_{1}\right\rangle
+\sin\left(  \frac{\theta}{2}\right)  e^{i\varphi}\left|  \phi_{2}%
\right\rangle
\]
where $\left|  \phi_{1}\right\rangle $ and $\left|
\phi_{2}\right\rangle $ are the eigenstates of $H$.
Recall that these states are not orthogonal but they define a complete basis on the $2\times 2$ space.

The general solution of the dynamics is of the form
\[
\left|  \Psi\left(  t\right)  \right\rangle
=c_{1}e^{-i\kappa_{1}t}\left|  \phi_{1}\right\rangle +c_{2}e^{-i\kappa_{2}%
t}\left|  \phi_{2}\right\rangle
\]
where $\kappa_{1}$ and $\kappa_{2}$ stand for the eigenvalues of $H$
corresponding to the mentioned eigenfunctions.
 One can adjust the coefficients  $c_{1}$ and $c_{2}$ in order to
 reproduce the chosen initial conditions. Then
\[
c_{1}=\cos\left(  \frac{\theta}{2}\right)\,\,\,\,;\,\,\,\,
c_{2}=\sin\left(  \frac{\theta}{2}\right)  e^{i\varphi}
\]
in an entirely similar way as for the hermitian case.

Now we determine the time $T$ taken by the system after the evolution to go back to the initial situation.
To do this, we write the state as
\[
\left|
\Psi(t)\right\rangle =e^{-\kappa_{1}i\,t}\left(  \cos\frac{\theta}{2}\left|
1\right\rangle +\sin\frac{\theta}{2}\,e^{i\varphi}e^{i\,t(\kappa_{1}%
-\kappa_{2})}\left|  2\right\rangle \right)
\]
and conclude that
\[
T\,(\kappa_{1}-\kappa_{2})=2\pi
\]
and that after this time $T$ the state acquires and extra phase because
\[
\left|  \Psi(T)\right\rangle
=e^{-i\Phi}\left|  \Psi\left(  0\right)  \right\rangle
\]
with
\[
\Phi
=2\pi\frac{\kappa_{1}}{(\kappa_{1}-\kappa_{2})}
\]

The next step is to extract from $\Phi$ the geometric phase. In order to do so we need to calculate the dynamical phase which is computed from an expression whose mathematical structure  is slightly different from the hermitian case due to the non orthogonality of the used base vectors \cite{Brody}, i.e.
\[
\Phi_{dynamic}=-\int_{0}^{T}\langle \tilde{\Psi}(t)|  H|  \Psi
(t)\rangle /\langle \tilde{\Psi}(t)|  \Psi
(t)\rangle dt
\]
where $ \langle \tilde{\Psi}(t)|$
 is the state but now expressed in the base of $H^{\dag}$, the hermitian  conjugate of $H$,
\[
\langle \tilde{\Psi}(t)|  =\tilde{c}_{1}e^{i\bar{\kappa}_{1}%
t}\langle \chi_{1}|  +\tilde{c}_{2}e^{i\bar{\kappa}_{2}t}\langle
\chi_{2}|
\]
and $ \langle \chi_{1}|  $ and
$\langle \chi_{2}|  $ are the members of the non-orthogonal base of $H^{\dag}$.
Since $\left\langle \chi_{n} |\phi
_{m}\right\rangle =\delta_{nm}$ $\left\langle \chi_{n}|
\phi_{n}\right\rangle $ and one can normalize the vector product $\langle \chi
_{n}|  \phi_{n}\rangle =1$ \cite{Brody} , then
\begin{eqnarray}
\langle \tilde{\Psi}(t)|  H|  \Psi(t)\rangle
&=&\tilde{c}_{1}c_{1}\kappa_{1}e^{i(\tilde{\kappa}_{1}-\kappa_{1})t}+\tilde{c}%
_{2}c_{2}\kappa_{2}e^{i(\tilde{\kappa}_{2}-\kappa_{2})t} \nonumber
\end{eqnarray}
Since we are in the region where the $PT$-symmetry is unbroken and consequently the eigenvalues are real,
one has
\[
\langle \tilde{\Psi}(t)|  H\left|  \Psi(t\right\rangle =\tilde
{c}_{1}c_{1}\kappa_{1}+\tilde{c}_{2}c_{2}\kappa_{2}=\cos^{2}\left(
\frac{\theta}{2}\right)  \kappa_{1}+\sin\left(  \frac{\theta}{2}\right).
\kappa_{2}
\]
Since
\[
\langle \tilde{\Psi}(t)|\Psi(t)\rangle=1,
\]
one concludes that we have recovered the result of the  the hermitian case. Consequently the dynamical phase  and the geometric phase are the same as in the hermitian case. This conclusion is only valid in the $\gamma$-parameter region where the $PT$-symmetry is valid.

\section{Classical analogs with resonant electric circuits}
We now revisit the geometric phase from the point of view of the classical analog Eqs.(\ref{coupled}) discussed previously.
We recall initially the analysis of the Hannay phase in the classical problem to proceed later on with the Berry phase in quantum systems with a  PT-symmetric non-hermitian Hamiltonian.

\subsection{Classical Foucault pendulum}
The differential equations of the electric system
equivalent to the Foucault pendulum includes a coupling by means of a gyrator \cite{AJP},
\begin{equation}
\left[
\begin{array}
[c]{c}%
\ddot{x}_{1}\\
\ddot{x}_{2}%
\end{array}
\right]  +\left[
\begin{array}
[c]{cc}%
0 & -a\\
a & 0
\end{array}
\right]  \left[
\begin{array}
[c]{c}%
\dot{x}_{1}\\
\dot{x}_{2}%
\end{array}
\right]  +\left[
\begin{array}
[c]{cc}%
\omega^{2} & 0\\
0 & \omega^{2}%
\end{array}
\right]  \left[
\begin{array}
[c]{c}%
x_{1}\\
x_{2}%
\end{array}
\right]  =0
\label{fouc}
\end{equation}
with $a=\gamma=2 \Omega\sin\lambda=\frac{G}{C}$ and $\omega^2=\frac{1}{LC}$. The procedure is to diagonalize the coupling term
to end with a pair of second degree uncoupled equations.

Diagonalizing the coupling term
\[
\left[
\begin{array}
[c]{cc}%
0 & -a\\
a & 0
\end{array}
\right],
\]
we obtain its eigenvalues $\pm i a$, and its eigenvectors that lead to the equivalent uncoupled equations
\begin{equation}
\left[
\begin{array}
[c]{c}%
\ddot{y}_{1}\\
\ddot{y}_{2}%
\end{array}
\right]  +i\left[
\begin{array}
[c]{cc}%
a & 0\\
0 & -a
\end{array}
\right]  \left[
\begin{array}
[c]{c}%
\dot{y}_{1}\\
\dot{y}_{2}%
\end{array}
\right]  +\left[
\begin{array}
[c]{cc}%
\omega^{2} & 0\\
0 & \omega^{2}%
\end{array}
\right]  \left[
\begin{array}
[c]{c}%
y_{1}\\
y_{2}%
\end{array}
\right]  =0.
\label{foucault}
\end{equation}
The new variables $y$ are related to the previous $x$ through
\[ \left[
\begin{array}
[c]{c}%
y_{1}\\
y_{2}%
\end{array}
\right]
=
\left[
\begin{array}
[c]{cc}%
1 & i\\
1 & -i
\end{array}
\right]  \left[
\begin{array}
[c]{c}%
x_{1}\\
x_{2}%
\end{array}
\right]  =\allowbreak\left[
\begin{array}
[c]{c}%
x_{1}+ix_{2}\\
x_{1}-ix_{2}%
\end{array}
\right] 
\]
The solution for the upper equation is  $y_{1}=e^{i\Lambda t}$.
Where the values of $\Lambda$, solutions of
$\Lambda^{2}+a\Lambda-\omega^{2}=0$ are
\[
\Lambda_{\pm}=\frac{1}{2}\,\left[  -a\pm\sqrt{a^{2}+4\omega^{2}}\right]
\]
under the standard approximation $\omega\gg a$ imposed by the physics of the considered systems, reduce to
\[
\Lambda_{\pm}%
\simeq-a/2\pm\sqrt{a^{2}+4\omega^{2}}/2=a/2\pm\omega
\]
Consequently, the general solution is
\[
y_{1}=Ae^{i\left(  a/2+\omega\right)
t}+Be^{i\left(  a/2-\omega\right)  t}
=e^{iat/2}\,\left(  A\,e^{-i\omega t}+B\,e^{i\omega t}\right)
\]

It is clear that  at $t=0$, $y_{1}\left(  0\right)  =A+B$, while at a time $T$ is
\[
y_{1}\left(
T\right)  =e^{iaT/2}\left(  A\,e^{-i\omega T}+B\,e^{i\omega T}\right)
\]
if $T$ is chosen as $T=2\pi/\Omega$ (remember the Foucault pendulum) and due to the fact that $\omega\gg\Omega$, one has
\[
\frac{\omega}{\Omega}=n\gg1
\]
This shows that the solution comes back to the initial state but in the evolution it
has acquired an extra phase \cite{AJP} 
\[
y_{1}\left(  T\right)  =e^{iaT/2}y_{1}\left(  0\right) =  e^{i\,\Omega\,\sin\lambda\,T/2}y_{1}\left(  0\right)
\]

\subsection{Classical analog of a simple quantum $PT$-symmetric model}

In the case of $PT$-symmetric systems, the experimental study in terms of electric circuits was pioneered in Ref. \cite{Schindler}.
We here consider the quantum $PT$-symmetric model 
as a classic counterpart defined by two simple resonant electric circuits but now coupled by a gyrator.
In this case the dynamical equations read
\begin{equation}
\left[
\begin{array}
[c]{c}%
\ddot{x}_{1}\\
\ddot{x}_{2}%
\end{array}
\right]  +\left[
\begin{array}
[c]{cc}%
-s & -g\\
g & s
\end{array}
\right]  \left[
\begin{array}
[c]{c}%
\dot{x}_{1}\\
\dot{x}_{2}%
\end{array}
\right]  +\left[
\begin{array}
[c]{cc}%
\omega^{2} & 0\\
0 & \omega^{2}%
\end{array}
\right]  \left[
\begin{array}
[c]{c}%
x_{1}\\
x_{2}%
\end{array}
\right]  =0
\label{nonher}
\end{equation}
As before we diagonalize the coupling term
\[
\left[
\begin{array}
[c]{cc}%
-s & -g\\
g & s
\end{array}
\right]
\]
to obtain the eigenvalues
\[
\Lambda_+=i\sqrt{\left(
-s^{2}+g^{2}\right)  }=i\tilde{a}\,\,\,\,;\,\,\,\,\Lambda_-= -i\sqrt{\left(  -s^{2}+g^{2}\right)  }=-i\tilde{a}
\]
where
\[
 \tilde{a}=\sqrt{\left(  g^{2}-s^{2}\right)  }=g\sqrt{\left(  1-\gamma^{2}\right)  }
 \]
where $\gamma=s/g$

In fact, we have reduced the equations to the previous hermitian case Eq.(\ref{foucault}) but now in terms of an effective $\tilde{a}$
that clearly goes into the previous situation in the limit $\gamma=0$

The solution $y_1$ now will get the original form in terms of $\tilde{a}$  but with the important detail that
in order to come back to the initial state the required time is not $T$ but a modified one
\[
\tilde{T}=\frac{2\pi}{\Omega\sqrt{\left( 1-\gamma^{2}\right) }}
\]
This modification implies that the acquired geometric phase is the same as
in the hermitian case, as it is easy to check from
\[
 \tilde{a}\tilde{T}/2 =\frac{1}{2}g\sqrt{\left(  1-\gamma^{2}\right)
}\frac{2\pi}{\Omega\sqrt{\left(  1-\gamma^{2}\right)  }}=\pi g/\Omega
\]

It is important to note that $\tilde{T}$ increases as soon as one is approaching, by modifying
the parameter $s$, the exceptional point where the $PT$-symmetry is spontaneously broken.

In summary, we have shown that the (classical) Hannay phase is related to the Berry phase of the equivalent quantum system,
in the case of $PT$ symmetry, as it was in the standard hermitian situation.

\begin{figure}[htb]
	  \includegraphics[scale=0.75]{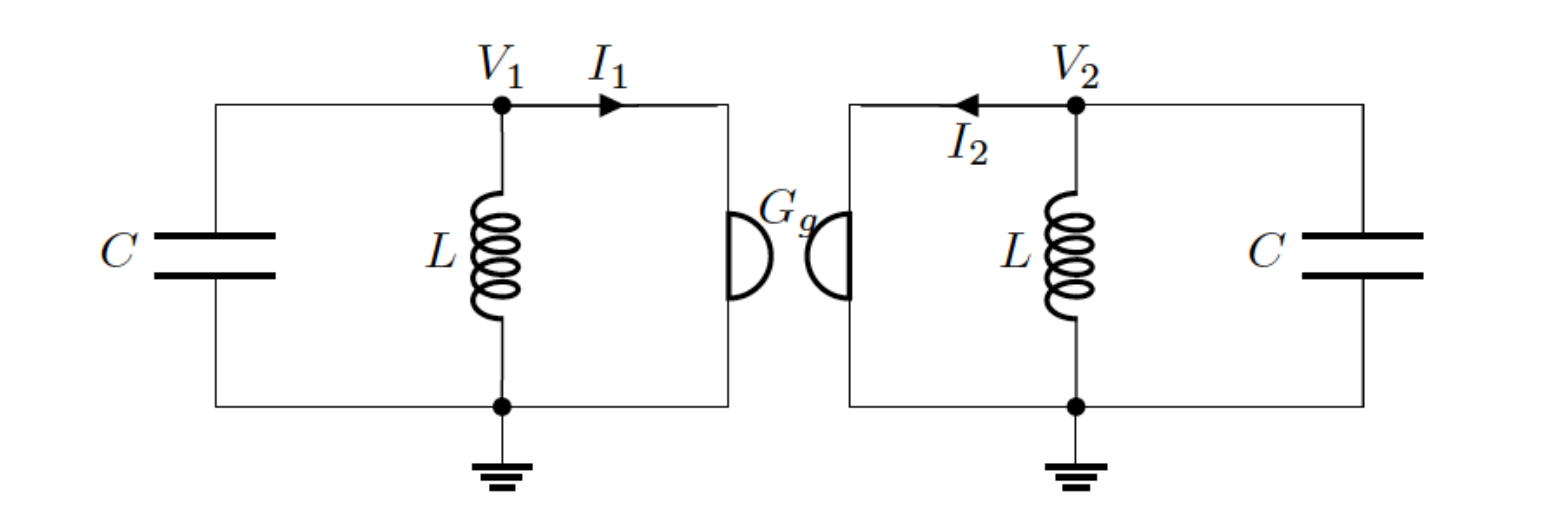}
\caption{Resonant circuits coupled by a Gyrator.}
	\label{circ_0}	
\end{figure}

\section{Resonant Electric Circuits}

The simulation of electric circuits with equations entirely identical to those of the examples previously presented gives a further insight on the appearance and behavior of the geometric phases as an holonomy effect.

\begin{figure}[htb]
         \includegraphics[scale=0.3]{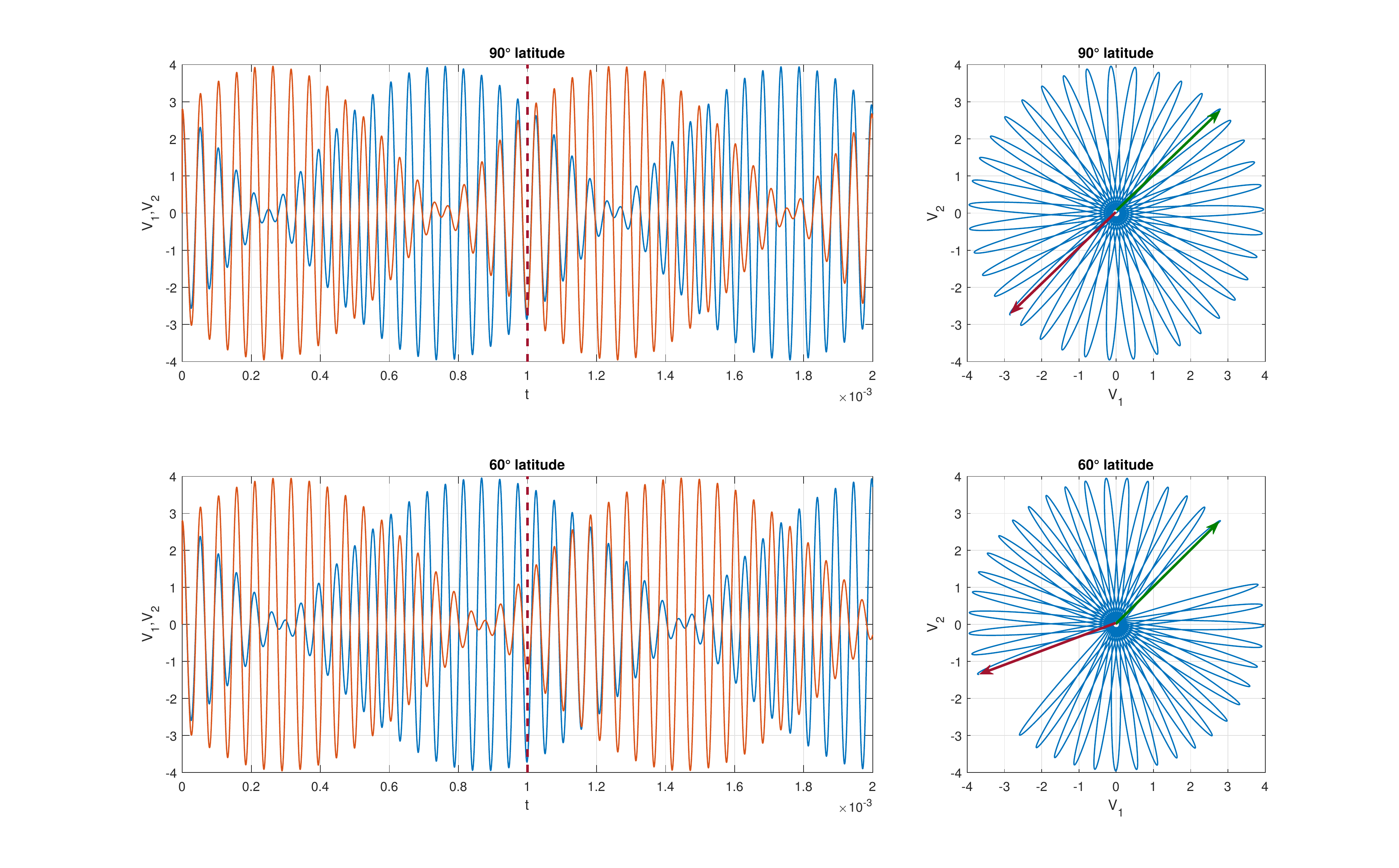}
	   	\caption{Simulation results for two different values of the gyrator (that correspond to different latitudes of the position of the Foucault pendulum). Given that initial conditions of both oscillators are equal, both plots start in the $45^{\circ}$ plane. The Hannay phase is the angle required to reach $2\pi$ after a period $T$ (the complement of the latitude effect on the pendulum) }
	\label{simus}
\end{figure}

We start with an electric network equivalent of the Foucault pendulum \cite{AJP} that is depicted in Fig. \ref{circ_0}. It consist of two identical, ideal $LC$ oscillators (without losses) coupled by a gyrator. The simple equations for voltages $V_1$ and $V_2$ are

\begin{equation}
\left\{
\begin{array}{r}
\ddot{V}_1+V_1\frac{1}{LC}-\dot{V}_2\frac{G_g}{C}=0\\
\ddot{V}_2+V_2\frac{1}{LC}+\dot{V}_1\frac{G_g}{C}=0\\
\end{array}
\right.
\end{equation}

Defining
 \[
 x_1=V_1\,\,\,;\,\,\,x_2=V_2\,\,\,;\,\,\,\frac{1}{LC}=\omega_0^2\,\,\,;\,\,\,\frac{G_g}{C}=a
  \]
  the expressions take exactly the form of Eq. (\ref{fouc}). In Fig. \ref{simus} the plot of $V_2$ vs. $V_1$ is depicted. In this figure the Hannay phase is 
  $2\,\pi$ minus the angle required to reach the initial position after an adiabatic closed excursion of period $T$.

The electronic simile of a $PT$-symmetric system based upon a pair of coupled $LC$ oscillators, one with loss (via $R$) and the other one with gain (via -$R$) allows the detection of the transition between a real spectrum of frequencies to a spontaneously broken $PT$-symmetry phase with complex frequencies \cite{Schindler}. This simple setup was performed with an inductive coupling between the oscillators, but it can be performed also when the coupling is through a gyrator, the non-reciprocal passive element of two ports with a given conductance \cite{Bala}. We notice that in this last case where a gyroscopic effect is present, the classical analog of a quantum $PT$-symmetric system reproduces all the known phenomena.

\begin{figure}[htb]
      \includegraphics[scale=0.75]{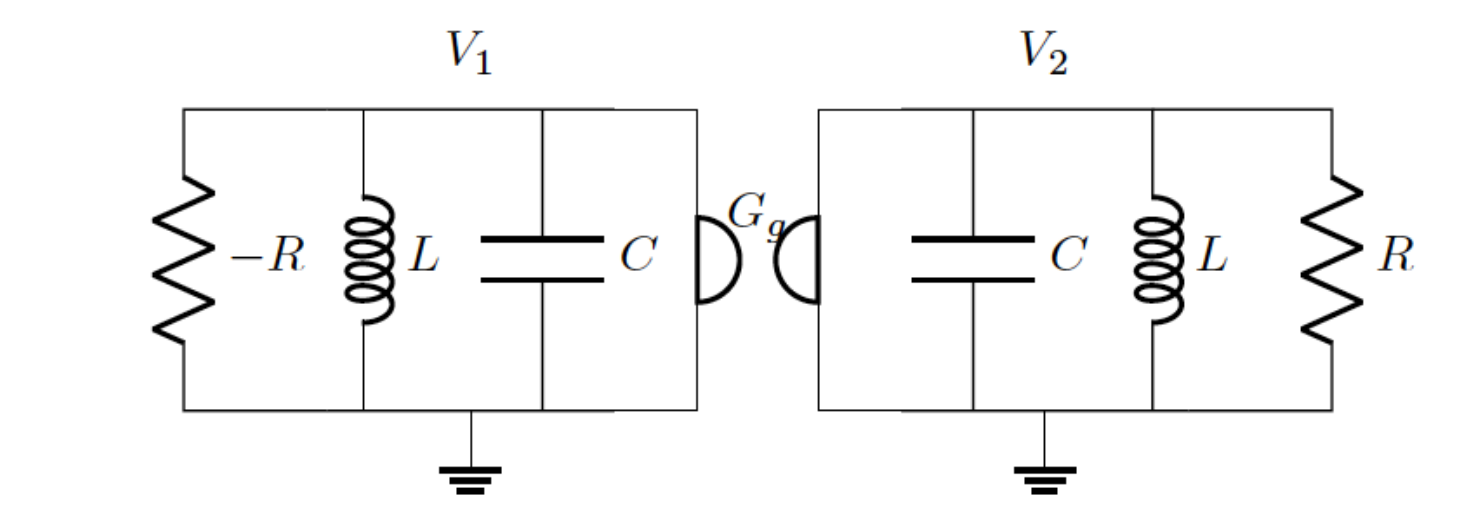}

\caption{Non-hermitian $PT$-symmetric circuit using a Gyrator}
\label{circ_1}
\end{figure}

The equivalent circuit to the $PT$-symmetric non-hermitic quantum system is shown in Fig. \ref{circ_1}. Resistor $R$ models the flow of energy outside the network, while $-R$ of equal absolute value, accounts for energy income and the system is $PT$-symmetric. The equations in this case are

\begin{equation}
\left\{
\begin{array}{r}
\ddot{V}_1+V_1\frac{1}{LC}-\dot{V}_1\frac{1}{RC}-\dot{V}_2\frac{G_g}{C}=0\\
\ddot{V}_2+V_2\frac{1}{LC}+\dot{V}_2\frac{1}{RC}+\dot{V}_1\frac{G_g}{C}=0\\
\end{array}
\right.
\end{equation}

\begin{figure}
\centering
 \includegraphics[width=10cm]{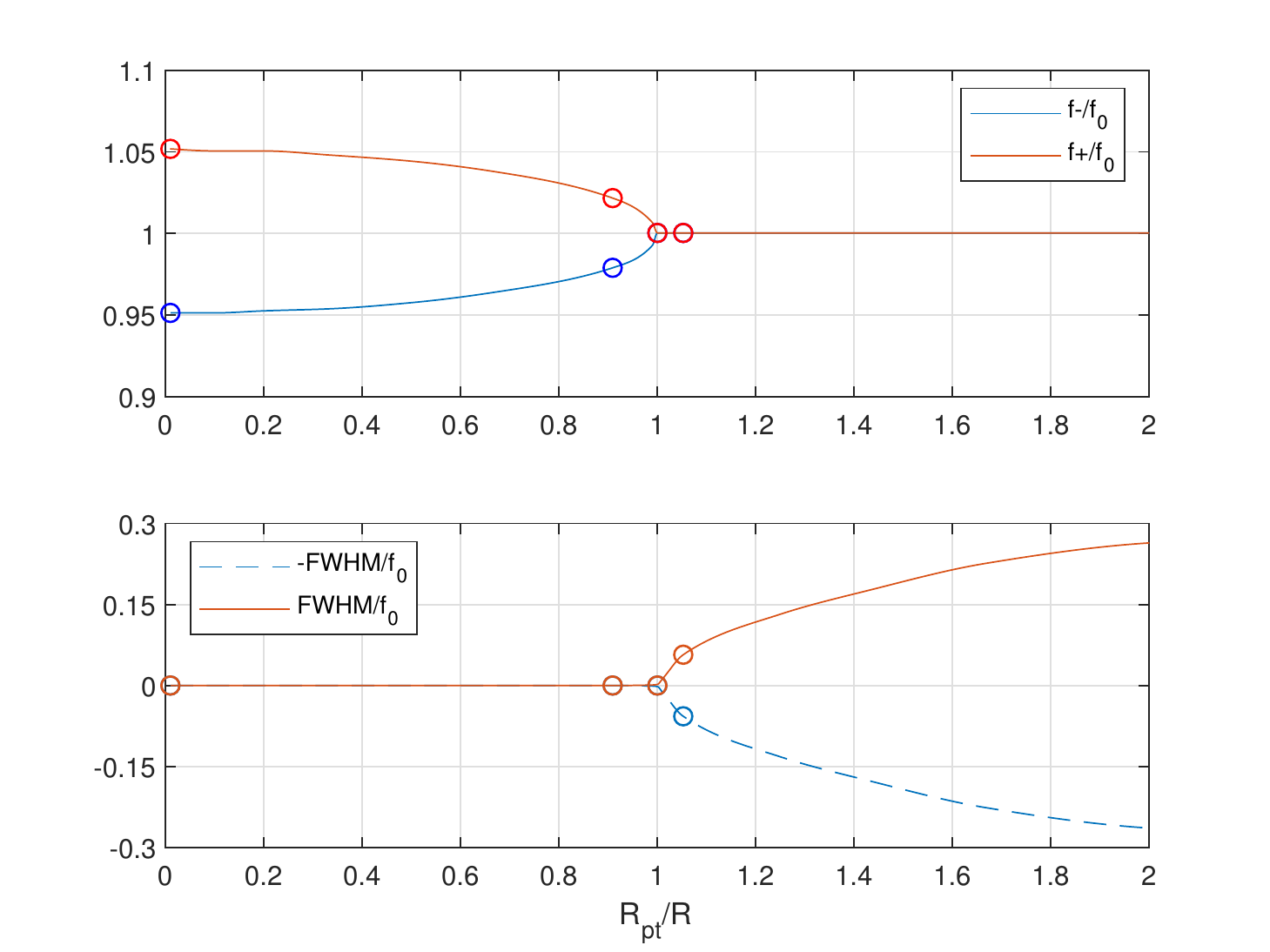}
\caption{Normalized imaginary an real parts vs. normalized degree of $PT$-symmetry for the non-hermitic circuit}
\label{simus2}
\end{figure}

Again, defining  
\[
x_1=V_1\,\,\,;\,\,\,x_2=V_2\,\,\,;\,\,\,\frac{1}{LC}=\omega_0^2\,\,\,;\,\,\,\frac{G_g}{C}=g\,\,\,;\,\,\,\frac{1}{RC}=s
 \]
 the equations take the form of Eq. (\ref{nonher}). The parameter $s$ accounts for the presence or not of the $PT$-symmetry. Fig. \ref{simus2} shows the evolution of the real and imaginary parts of the eigenvalues as $s$ changes, describing the path to an exceptional point, where the spontaneous loss of $PT$-symmetry occurs and the eigenvalues become complex conjugate. In this way our laboratory simulation shows explicitly the occurrence of a phase transition in this non hermitian system from a phase in which the $PT$-symmetry is realized to a phase in which it is spontaneously broken.

\section{Conclusions}

Both, the well-known Berry phase, the geometric phase present in the quantum adiabatic evolution and  its classical analogue, the Hannay phase
where computed and analyzed in the simple case of a finite dimensional dynamics. The classical and quantum geometric phases for equivalent cases are related through a very precise numerical factor that was previously predicted \cite{Berry1}. The study was also done for the case of hermitian and non-hermitian but $PT$ symmetric quantum Hamiltonians and the corresponding classical mathematically equivalent dynamics. It is worth mentioning that in the region where the $PT$-symmetry is present in the solution, the geometric phase agrees with the one obtained in the  hermitian case.

Taking profit of the {\it stricto sensu} equivalence between classical and quantum dynamics \cite{AP} was possible to build up resonant electric circuits, coupled by means of gyrators, that reproduce exactly the theoretical solutions obtained. In this way one is able to show explicitly a quantum mechanical phase transition from a phase in which a symmetry is realized to a phase in which it is spontaneously broken. Moreover, our construction allows one to discriminate between couplings and to show that the gyrator coupling is favored not only by its simpler analysis but also for pedagogical reasons, because, as it was previously shown \cite{AJP,M}, in this case the systems can be put in one to one correspondence with the Foucault pendulum.

\section*{Acknowledgements}
HF and CAGC were partially supported by ANPCyT, Argentina. VV was
supported by MCIN/AEI/10.13039/501100011033, European Regional Development Fund Grant No. PID2019-105439 GB-C21 and by GVA PROMETEO/2021/083 .

\section*{Appendix}

As mentioned in the main text, the $PT$-symmetry is spontaneously broken in terms of a  parameter $\gamma$
when it goes through an exceptional point.

As it is well known an operator, as $PT$, represents a symmetry if it commutes with the Hamiltonian of the system, namely
\[
 \left[ H,PT\right] =0
 \]
In the case of  $PT$, the Hamiltonian has real eigenvalues below the exceptional point $\gamma^{\star}$
and complex conjugate ones above it. Explicitly
\[
H\Psi _{1,2}=\lambda _{1,2}\Psi _{1,2}
\]
with $\lambda _{1}$ and $\lambda _{2}$
real for $\gamma < \gamma^{\star}$ and with complex $\lambda _{2}=\lambda _{1}^{\ast }$ for $\gamma > \gamma^{\star}$

The presence of the symmetry implies that
\[
H\left( PT\right) \Psi _{1,2}=\left(
PT\right) \left( \lambda _{1,2}\Psi _{1,2}\right)
\]
and in the region where the eigenvalues are real one has
\[
PT\left( \lambda _{1,2}\Psi
_{1,2}\right) =\lambda _{1,2}PT\Psi _{1,2}
\]
meaning that $PT\Psi _{1,2}$ is also eigenfunction of $H$. Then $\Psi _{1,2}$, the solution, has
the same symmetry present in the Hamiltonian. In this region the symmetry has a Wigner-Weyl  realization.

Above the  exceptional point the situation is different because the symmetry operator also acts on the complex eigenvalues
conjugating them.
 \[
H\left( PT\right) \Psi _{1}=\left( PT\right) \left( \lambda _{1}\Psi
_{1}\right) =\lambda _{1}^{\ast }\left( PT\right) \Psi _{1}=\lambda
_{2}\left( PT\right) \Psi _{1}
\]
or
\[
\left( PT\right) \Psi _{1}=\Psi_{2}
\]
which explicitly shows that the $PT$-symmetry is not present in the solution and for this reason the
spontaneous symmetry breaking appears. The symmetry of $H$ has a Nambu-Goldstone realization . Notice that we are
dealing with a discrete symmetry and therefore no Goldstone boson appears.

\end{document}